\documentclass[10pt,showpacs,twocolumn]{revtex4}
\usepackage{graphicx}
\usepackage{amsmath}
\usepackage{amssymb}
\usepackage{}

\begin{document}
\title{Asymmetric electron energy sharing in strong-field double ionization of helium}
\author{Yueming Zhou, Qing Liao and Peixiang Lu \footnote{Corresponding author: lupeixiang@mail.hust.edu.cn}}

\affiliation{Wuhan National Laboratory for Optoelectronics, Huazhong
University of Science and Technology, Wuhan 430074, P. R. China}
\date{\today}

\begin{abstract}
With the classical three-dimensional ensemble model, we have
investigated the microscopic recollision dynamics in nonsequential
double ionization of helium by 800 nm laser pulses at 2.0 PW$/cm^2$.
We demonstrate that the asymmetric energy sharing between the two
electrons at recollision plays a decisive role in forming the
experimentally observed V-shaped structure in the correlated
longitudinal electron momentum spectrum at the high laser intensity
[Phys. Rev. Lett. {\bf 99}, 263003 (2007)]. This asymmetric energy
sharing recollision leaves footprints on the transverse electron
momentum spectra, which provide a new insight into the attosecond
three-body interactions.

\end{abstract}  \maketitle

Nonsequential double ionization (NSDI) of atom in strong laser field
has drawn extensive researches in the recent years because it
provides a particular clear manner to study the electron-electron
correlation, which is responsible for the structure and the
evolution of large parts of our macroscopic world
\cite{Weber,Becker}. The measurements of the recoil ion momentum
distributions \cite{Moshammer,Rudenko}, the electron energy
distributions \cite{Lafon,Chaloupka}, the correlated two-electron
momentum spectra \cite{Feuerstein,Moshammer2}, as well as numerous
theoretical calculations \cite{Lein,Becker2,Ho,Haan} have provided
convincing evidences that strong-field NSDI occurs in favor of the
classical recollision model \cite{Corkum}. According to this model,
the first electron that tunnels out of the atom picks up energy from
the laser field, and is driven back to its parent ion when the field
reverses its direction, and transfers part of its energy to dislodge
a second electron. Though the recollision model describes the NSDI
process in a clear way, the details of recollision remain obscure.
For instance, at intensities below the recollision threshold, the
underlying dynamics for the intensity-independent 5U$_p$ (U$_p$ is
the ponderomotive energy) cutoff in the two-electron energy spectra
\cite{Parker,Liu,Zhou} and the dominant back-to-back emission of the
correlated electrons from NSDI of Ar \cite{Liu} has not been well
explored.

Recently, the high resolution and high statistics experiments on
double ionization (DI) of helium have made a great progress in
unveiling the microscopic recollision dynamics in NSDI. The
finger-like structure in the correlated longitudinal (in the
direction parallel to the laser polarization) momentum distribution
from NSDI of helium by a 800 nm, 4.5 $\times 10^{14}$ W$/cm^2$ laser
pulses indicates backscattering at the nucleus upon recollision
\cite{Staudte}. At a higher intensity, 1.5 $\times 10^{15}$
W$/cm^2$, Rudenko $et$ $al$ observed a pronounced V-like shape of
the correlated two-electron momentum distribution \cite{Rudenko2},
which is interpreted as a consequence of Coulomb repulsion and
typical (e,2e) kinematics. Theoretical studies have demonstrated
that at the relatively low laser intensity, both the nuclear Coulomb
attraction \cite{Haan2,Ye} and the final-state electron repulsion
\cite{Ye,Chen} contribute to this novel structure. However, at the
relatively high laser intensity, the roles of final-state electron
repulsion and nuclear attraction for the V-like shape have not been
examined. It is questionable whether the responsible microscopic
dynamics for the V-like shape at this high intensity is similar to
that at the relatively low intensity.

In this Letter, with the classical three-dimensional (3D) ensemble
model \cite{Haan,Zhou2}, we examine the microscopic recollision
dynamics in NSDI of helium by a high intensity (2.0$\times 10^{15}$
W$/cm^2$) laser pulse. We find that the pronounced V-like shape of
the correlated electron momentum in the direction parallel to the
laser polarization is a consequence of the asymmetric electron
energy sharing in the recollision process, whereas neither the
nuclear attraction nor the final-state electron repulsion
contributes to the V-like shape. This is different from that at
relatively low intensity, where both the nuclear Coulomb attraction
and final-state electron repulsion play significant roles in forming
the finger-like shape. By separating the recolliding electron from
the bound electron, we find that the transverse (in the direction
perpendicular to the laser polarization) momentum spectra for these
two groups of electrons peak at different momenta. This difference
is ascribed to the Coulomb focusing in the transverse direction when
the electron moves away from the core and can be understood as a
footprint of the asymmetric electron energy sharing at recollision.

The 3D classical ensemble model is introduced in \cite{Haan} and
widely recognized as an useful approach in studying high-field
double ionization. In this classical model, the evolution of the
two-electron system is governed by the Newton's classical equations
of motion (atomic units are used throughout this Letter unless
stated otherwise): $\frac{d^2 {\bf r}_i}{dt^2}=-\nabla
[V_{ne}(r_i)+V_{ee}(r_1,r_2)]-{\bf E}(t)$, where the subscript $i$
is the label of the two electrons, and ${\bf E}(t)$ is the electric
field, which is linearly polarized along the x axis and has a
trapezoidal pulse shape with four-cycle turn on, six cycles at full
strength, and four-cycle turn off. The potentials are
$V_{ne}(r_i)=-2/\sqrt{r_i^2+a}$ and
$V_{ee}(r_1,r_2)=1/\sqrt{(r_1-r_2)^2+b}$, representing the
ion-electron and electron-electron interactions, respectively. The
soft parameter a is set to 0.75 to avoid autoionization and b is set
to 0.01 \cite{Haan,Zhou2}. To obtain the initial value, the ensemble
is populated starting from a classically allowed position for the
helium ground-state energy of -2.9035 a.u. The available kinetic
energy is distributed between the two electrons randomly in momentum
space, and then the electrons are allowed to evolve a sufficient
long time in the absence of the laser field to obtain stable
position and momentum distributions \cite{Zhou}. Note that in the
classical model the first electrons are ionized above the suppressed
barrier and no tunneling ionization occurs.

Figures 1(a) and 1(b) display the correlated electron momentum
distributions in the direction parallel to the laser polarization,
where the laser intensities are 5.0 $\times 10^{14}$ W$/cm^2$ and
2.0 $\times 10^{15}$ W$/cm^2$, respectively. At 5.0 $\times 10^{14}$
W$/cm^2$, the experimental observed finger-like structure is not
reproduced (Fig. 1(a)). This is because of the large soft parameter
employed in our calculation, which shields the nuclear potential
seriously. Previous studies have illustrated that the finger-like
structure is able to be reproduced when the realistic Coulomb
potential or a soften potential with a smaller screening parameter
is used \cite{Haan2,Ye}.

At the relatively high intensity, the overall V-like shape in the
correlated momentum distribution is obvious. In contrast to the
previous experimental result \cite{Rudenko2}, a cluster of
distribution around zero momentum is clearly seen. Back analysis
reveals that these events correspond to the trajectories where DIs
occur at the turn-on stage of the laser pulse. For the soft
potential employed in this Letter, the potential energy well for the
second electron is $-2/\sqrt{0.75}\simeq-2.3$ a.u., which is lower
than that of realistic helium. In the classical description, the
first electron can get ionized more easily at the expense of leaving
the second electron near the bottom of the potential well
\cite{Haan3}. Thus the first electron can be ionized very early at
the turn-on stage of the pulse, leading to recollision occurs at the
turn-on stage. This effect results in an overestimated contribution
from the turn-on stage of the laser pulse to DI. In order to
overcome this deficiency and focus our study on the high intensity
regime, we artificially exclude the events in which DI occurs at the
turn-on stage of the laser pulse, as shown in Fig. 1(c). The
correlated electron momentum distribution agrees excellently well
with the experiment \cite{Rudenko2} and the V-like shape is obvious
though a soft parameter as large as a=0.75 is employed. We also
performed further calculations by changing the soft parameter $a$
after the first ionization \cite{Zhou,Haan2}, and no noticeable
change has been found in the V-like shape. It implies that the
nuclear attraction does not contribute to the V-like shape, which is
different from that at the relatively low laser intensity
\cite{Haan2,Ye}.

It has been confirmed that at the relatively low intensity, the
final-state electron repulsion plays an important role for the
finger-like shape of the correlated electron momentum distribution
\cite{Ye,Chen}. In order to examine the role of final-state electron
repulsion in forming the V-shape at the high intensity, we have
performed an additional calculation, in which the final-state
electron interaction $V_{ee}(r_1,r_2)=1/\sqrt{(r_1-r_2)^2+b}$ is
replaced by $V_{ee}(r_1,r_2)=exp{[-\lambda r_b]}/r_b$, where
$r_b=\sqrt{(r_1-r_2)^2+b}$ and $\lambda=5.0$ \cite{Ye}. As shown in
Fig. 1(d), the V-like shape is still clearly seen, and no noticeable
difference is found when compared to Fig. 1(c). Thus it confirms
that the V-like shape is not a consequence of the final-state
electron repulsion at this high intensity.

The analysis above illustrates that neither the nuclear attraction
nor the final-state electron repulsion contributes to the V-like
shape in the correlated longitudinal electron momentum at the high
laser intensity. In order to explore the responsible dynamics for
the V-like shape, we take further advantage of back analysis
\cite{Ho}. Tracing the classical DI trajectories allows us easily to
determine the recollision time and the energy exchange during
recollision. Here, the recollision time is defined to be the instant
of the closest approach after the first department of one electron
from the core \cite{Haan}.

In Figs. 2(a) and 2(b), we have segregated the trajectories shown in
Fig. 1(c) according to the energy difference of the two electrons at
time 0.02T after recollision (T is the laser period. We have changed
the time from 0.02T to 0.05T and the conclusions below do not change
with the variation of this time.). Figures 2(a) and 2(b) display the
correlated longitudinal electron momentum distributions of the
trajectories where the energy difference is larger and less than 2.0
a.u., respectively. It is clearly shown that the events are
clustered on the main diagonal when the two electrons achieve
similar energies at recollision (Fig. 2(b)). In contrast, the
correlated electron momentum distribution exhibits distinct
off-diagonal features when asymmetric energy sharing (AES) occurs
(Fig. 2(a)). Based on these results, we can conclude that the AES at
recollision is the decisive reason for the V-like shape in the
longitudinal electron momentum correlation at the high laser
intensity.

In order to further understand the AES at this high laser intensity,
in Figs. 2(c) and 2(d) we present the counts of DI trajectories
versus laser phase at recollision. Figures 2(c) and 2(d) correspond
to the trajectories from Figs. 2(a) and 2(b), respectively. It is
found that in Fig. 2(c), where AES occurs, recollisions cluster
around the zero crossing of the laser field, while in Fig. 2(d),
recollisions occur close to the extremum of the field. According to
the simple-man model \cite{Corkum}, the electrons with the maximal
recolliding energy return to the core near the zero crossing of the
laser field. While these returning to the core near the extremum of
the field possess lower recolliding energies. Figures 2(c) and 2(d)
imply that the energetic recollisions often favor AES while the less
energetic ones tend to have more symmetric energy sharing (SES).
After distinguishing the recolliding electrons from the bound
electrons we find that for 88\% of the AES events (the events in
Fig. 2(a)) the energy of the recolliding electron just after
recollision is higher than that of the bound electron. It indicates
that in the high returning-energy recollision, the recolliding
electron only transfers a small part of its energy to the bound
electron. This issue is consistent with a recent study
\cite{Mauger}, in which it has been demonstrated that the efficacy
of electron-electron collisions decreases with the increasing
collision energy.

At the relatively low laser intensity, because of the lower
recolliding energy, the efficacy of energy exchange at recollision
is high. Thus AES is not serious and its contribution to the
finger-like structure is negligible. At the high laser intensity,
the low energy exchange efficacy at recollision makes AES play the
dominant role in forming the V-like shape in the correlated electron
momentum spectrum. Because of the dramatic AES, the two electrons
leave the core with very different initial momenta and separate
quickly. As a consequence, the final-state electron repulsion is
weak and does not contribute to the V-like shape.

More details of recollision can be obtained by inspecting the
transverse momenta because the subtleties of the momentum exchange
in the recollision process are not covered by the much larger
momentum transfer taken from the laser field \cite{Wechenbrock}. In
Figs. 3(a) and 3(b), we present the joint-probability distributions
of the transverse momenta (along y axis) for the events shown in
Figs. 2(a) and 2(b), respectively. Remarkably, in Fig. 3(b) the
distribution lies along the diagonal $p_{1y}+p_{2y}=0$. This
behavior indicates the strong repulsion in the transverse direction,
which is in agreement with precious studies \cite{Wechenbrock}.
Contrarily, in Fig. 3(a) the population is clustered along the axes
$p_{1y}=0$ and $p_{2y}=0$, indicating different amplitudes of
transverse momenta of the two electrons. This difference is more
clear when separating the bound electrons from the recolliding ones.
In the bottom of Fig. 3, we display the transverse momentum
($P_{i\bot}=\sqrt{p_{iy}^2+p_{iz}^2}$) spectra of the recolliding
(red circle) and the bound (black triangle) electrons separately,
where Figs. 3(c) and 3(d) correspond to the events from Figs. 3(a)
and 3(b), respectively. For the SES trajectories (Fig. 3(d)), the
recolliding and the bound electrons exhibit similar transverse
momentum distributions. Whereas for the AES ones (Fig. 3(c)), the
difference in the distributions of the recolliding and bound
electrons is remarkable: the spectrum of the bound electrons peaks
near 0.2 a.u., while for the recolliding electrons the spectrum
exhibits a maximum at 1.2 a.u. The different transverse momentum
distributions for the SES and AES trajectories imply the different
three-body interactions, which can be explored by monitoring the
history of the DI events.

We display two sample trajectories in Fig. 4. In the left column,
the two electrons have equal energy after recollision (Fig. 4(a)),
and achieve similar final longitudinal momentum (Fig. 4(c)). For the
trajectory shown in the right column, the two electrons share
unequal energies upon recollision. The recolliding electron (solid
red curve) obtains a higher energy at recollision (Fig. 4(b)) but
achieves a smaller final longitudinal momentum (Fig. 4(d)) due to
the postcollision velocity \cite{Haan}. The time evolution of the
transverse momentum is more interesting. As shown in the bottom of
Fig. 4, for both trajectories the two electrons obtain similar
transverse momenta with opposite directions upon recollision. For
the SES trajectories, both electrons experience a small sudden
decrease in the transverse momenta just after recollision (Fig.
4(e)). For the AES trajectory, the bound electron suffers a much
larger sudden decrease in the transverse momentum while the
transverse momentum of the recolliding electron does not change
after recollision (Fig. 4(f)). We ascribe the sudden decrease of the
transverse momentum to the nuclear attraction in the transverse
direction when the electron moves away from the core. For the SES
trajectories, the two electrons leave the core with similar
momentum, thus the nuclear attraction plays a similar role in
decreasing the transverse momentum, resulting in the distribution
along the diagonal $p_{1y}+p_{2y}=0$ in Fig. 3(b). For the AES
trajectory, the nucleus does not effect the transverse momentum of
the recolliding electron because it leaves the core with a very fast
initial momentum. While for the bound electron, it takes a longer
time to leave the effective area of the core due to the small
initial momentum, leading to a significant decrease of the
transverse momentum caused by nuclear attraction. The transverse
momentum change of the electron is determined by $\Delta
p_{\bot}=\int F_{\bot}dt$, where $F_{\bot}$ is the transverse force
of the nuclear attraction. Assuming an electron that starts at a
field zero near the region x=2 a.u. with initial momentum
$\upsilon_{\bot}=1.2$ a.u. and evolves in the combined laser and
Coulombic field, it takes a time about 10 a.u. for the nucleus to
decrease $\upsilon_{\bot}$ to 0.2 a.u.

Simply speaking, in the AES trajectory, because of the different
initial momentum, the nuclear attraction plays different roles in
``focusing" the transverse momenta of the bound and recolliding
electrons, resulting in the momentum distributions in Fig. 3(c). In
other words, the different transverse momentum distributions of the
recolliding and bound electrons reflect the AES at recollision and
provide a new insight into the attosecond three-body interactions.

In conclusion, we have investigated the attosecond recollision
dynamics in NSDI of helium at 2.0 $\times 10^{15}$ W/$cm^2$. At the
high intensity, the bound electron often shares a small part of the
recolliding energy at recollision due to the low efficacy of energy
exchange at the high recolliding energy. This asymmetric energy
sharing is the decisive reason for the observed V-like shape in the
correlated longitudinal momentum spectrum at the high laser
intensity. Because of the asymmetric energy sharing recollision, the
bound electron leaves the core with a small initial momentum. Thus
its transverse momentum is strongly focused by the nuclear
attraction when it moves away from the core. Whereas the recolliding
electron leaves the core so fast that its transverse momentum is not
effected by the nuclear attraction. The different transverse
momentum spectra of the recolliding and bound electrons act as a
signature of the asymmetric energy sharing at recollision and
provide a new insight into the attosecond three-body dynamics.

This work was supported by the National Natural Science Foundation
of China under Grant No. 10774054, National Science Fund for
Distinguished Young Scholars under Grant No.60925021, and the 973
Program of China under Grant No. 2006CB806006.

\newpage
\begin{figure}[p]
\begin{center}
\includegraphics[width=8.6cm,clip]{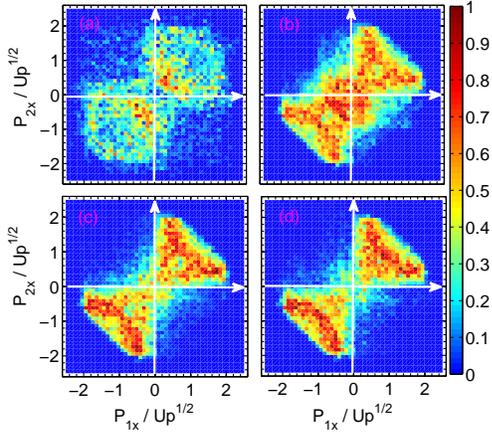}
\caption{\label{fig1} (color online) Correlated longitudinal
electron momentum distributions for NSDI of helium by 800 nm laser
pulses. The intensities are (a) 0.5 PW/$cm^2$ and (b)-(d) 2.0
PW/$cm^2$.  In (c) and (d), the trajectories where DI occurs at the
turn-on stage of the trapezoidal pulse are excluded. In (d), the
final-state e-e repulsion is neglected by replacing the soft Coulomb
repulsion with Yukawa potential (see text for detail). The ensemble
sizes are 2 millions.}
\end{center}
\end{figure}

\begin{figure}[p]
\begin{center}
\includegraphics[width=8.6cm,clip]{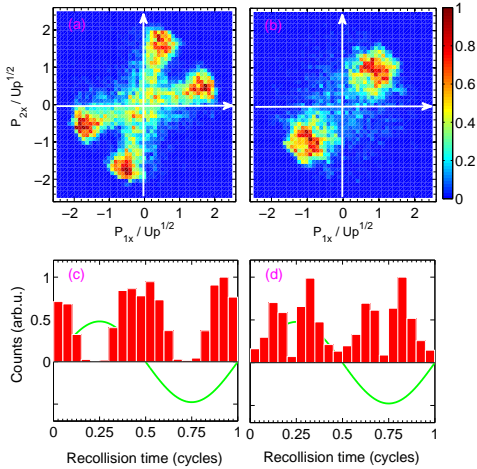}
\caption{\label{fig2}(color online) Correlated longitudinal electron
momentum distributions for the trajectories where the energy
difference at time 0.02T after recollision is (a) larger than 2 a.u.
and (b) smaller than 2 a.u. (c)(d) Counts of DI trajectories versus
laser phase at recollision for the events in (a) and (b),
respectively. The solid green curves represent laser fields. In all
plots, the events where DI occurs at the turn-on stage of the
trapezoidal pulse are excluded.}
\end{center}
\end{figure}

\begin{figure}[p]
\begin{center}
\includegraphics[width=8.6cm,clip]{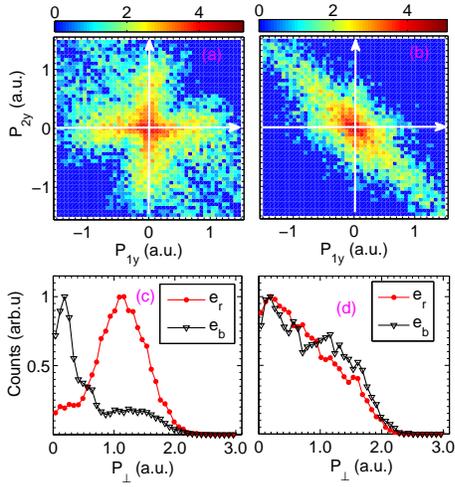}
\caption{\label{fig3}(color online) (a)(b) Joint-probability
distributions (on logarithmic scale) of the transverse momenta
(along y axis) for the trajectories from Figs. 2(a) and 2(b),
respectively. (c) Transverse momentum spectra of recolliding (red
cycle) and bound (black triangle) electrons for the trajectories
from (a). (d) The same as (c) but for the trajectories from (b).}
\end{center}
\end{figure}

\begin{figure}[p]
\begin{center}
\includegraphics[width=8.6cm,clip]{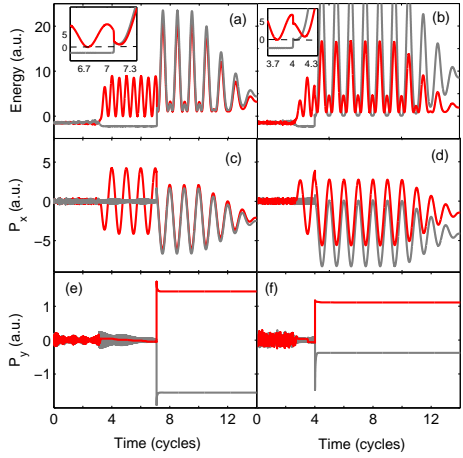}
\caption{\label{fig4} (color online) Two sample trajectories
selected from Fig. 2(a) (right column) and Fig. 2(b) (left column),
respectively. The upper, middle and bottom rows show the energy,
longitudinal momentum and transverse momentum (along y axis) versus
time for each electron, respectively. The energy exchange at
recollision is clearly visible in the insets of (a) and (b).}
\end{center}
\end{figure}

\end{document}